\documentclass[twocolumn,showpacs,preprintnumbers,amsmath,amssymb]{revtex4}

\usepackage{graphicx}
\usepackage{dcolumn}
\usepackage{bm}

\begin{document}


\title{Novel Complete Non-compact Symmetries for the Wheeler-DeWitt Equation in a  Wormhole Scalar Model and  Axion-Dilaton String Cosmology}

\author{Rub\'en Cordero}
\email{cordero@esfm.ipn.mx}
\affiliation
{
Departamento de F\'\i sica, Escuela Superior de F\'\i sica y Matem\'aticas del
I. P. N., \\
 Unidad Profesional Adolfo L\'opez Mateos, Edificio 9, 07738 M\'exico D.F., M\'exico.}
\author{Victor D. Granados}
\email{granados@esfm.ipn.mx}
\affiliation{Departamento de F\'\i sica, Escuela Superior de F\'\i sica y Matem\'aticas del
I. P. N., \\
Unidad Profesional Adolfo L\'opez Mateos, Edificio 9, 07738 M\'exico D.F., M\'exico.}
\author{Roberto D. Mota}
 \email{rmotae@ipn.mx}
\affiliation{ Departamento de ICE de la Escuela Superior de
Ingenier\'\i a Mec\'anica y El\'ectrica del I. P. N., Unidad
Culhuacan. Av. Santa Ana No. 1000, San Francisco Culhuacan, Coyoacan
M\'exico D. F.,\\ C. P. 04430, M\'exico.}

%
\date{\today}

\begin{abstract}
We find the full symmetries of the Wheeler-DeWitt equation for the
Hawking and Page wormhole model and an axion-dilaton string
cosmology. We show that the Wheeler-DeWitt Hamiltonian admits an
$U(1,1)$ hidden symmetry for the Hawking and Page model and $U(2,1)$
for the axion-dilaton string cosmology. If we consider the existence
of matter-energy renormalization, for each of these models we find
that the Wheeler-DeWitt Hamiltonian accept an additional $SL(2,R)$
dynamical symmetry. In this case, we show that the $SL(2,R)$
dynamical symmetry generators transform the states from one energy
Hilbert eigensubspace to another. Some new wormhole type-solutions
for both models are found.
\end{abstract}

\maketitle

\section{Introduction}

The study of the early universe has become one of the more intense
research areas in physics. General relativity describes the
universe at scales larger than the Planck scale and it is expected
that quantum mechanics has to be taken into account at least at
these small scales.

In the quantum cosmology framework the whole universe is
represented by means of a wave function. The quantum cosmology
formalism, including the definition of the wavefunction of the
universe, its configuration space  and its evolution according to
the Wheeler-DeWitt equation, was set up in the late 1960s
\cite{1,2,3,4,5}.

The development of quantum cosmology started at the beginning of
1980 when was proposed that the universe could be spontaneously
nucleated out of nothing \cite{7}, where nothing means the absence
of space and time. After nucleation the universe enters to a phase
of inflationary expansion and continues its evolution to the
present. However there are several important questions that remain
to be solved like the appropriate boundary conditions for the
Wheeler-DeWitt equation. In the case of quantum mechanics there is
an external setup and the boundary conditions can be imposed
safely, but in 4-dimensional quantum cosmology there is nothing
external to the universe and the correct boundary condition
remains unsolved. A number of proposals for such boundary
conditions came out \cite{8,9,10,11,12}. Once we have chosen the
boundary conditions, different physical settings emerge, e. g. if
the wave function is regular when the three-geometry collapses to
zero and it is exponentially damped for large three-geometries we
have wormholes \cite{HP}. Hawking and Page \cite{HP} considered a
minimally coupled massless scalar field and found its wormhole
solution.

On the other hand, it is believed that string theory could play an
important role in describing the evolution of the early universe
and gives some insights on the mechanism of inflation. The pre-big
bang scenario \cite{PBB} is very interesting since it uses stringy
symmetries in order to give a novel mechanism for inflation.
Besides, it is possible that string theory could provide
resolution of the initial singularity problem in cosmology.

In string cosmology the usual approach is to study time-dependent
solutions to the lowest-order string equations of motion. This
standard approach applies on scales above those energies where the
string symmetries are broken but on scales below the string scales
\cite{LIDSEYREP}. The low energy four dimensional effective field
theory action of string theory  contains two massless fields
\cite{st}. One of these scalars fields is called the axion $\chi$
and it comes from the third rank field strength corresponding to
the Kalb-Ramond field, the other one is called the dilaton $\phi$.
The physical consequences of the axion in a curved spacetime has
been investigated in the aim of finding possible indirect
evidences of low energy string theory \cite{axion1, axion2,
axion3}. The dilaton is very important in string theory since it
defines the string coupling constant $g_s$  as $e^{\phi/2}$, it
determines the Newton constant, the gauge coupling constants and
Yukawa couplings.

It is a well known fact that symmetries are very important to
understand several properties of diverse theories. In particular,
it is very interesting to investigate the underlying symmetries of
the Hawking and Page wormhole model and axion-dilaton string
cosmology. For the first model, a $U(1)$ symmetry generated by the
``angular momentum'' is present.  For the second model, Maharana
\cite{maha1,maha2} showed that the ``angular part'' of the
Wheeler-DeWitt equation is invariant under the $SO(2,1)$ group of
transformations. For both models the angular symmetries were
employed to reduce the Wheeler-DeWitt equation to a
one-dimensional radial equation. However, in this work we show
that these systems have larger symmetry groups.

In this paper we consider the axion-dilaton string cosmology studied
by Maharana and find the $U(2,1)$ symmetry for the complete
Wheeler-DeWitt Hamiltonian. For the Hawking and Page wormhole model
we find that the Wheeler-DeWitt Hamiltonian admits an $U(1,1)$
symmetry. Also, for each of these models we show that the
Wheeler-DeWitt Hamiltonian accept an additional  $SL(2,R)$ dynamical
symmetry when matter-energy renormalization is allowed. In this
case, we prove that the $SL(2,R)$ dynamical symmetry transform
states from an energy Hilbert eigensubspace to another energy
eigensubspace. The paper is organized as follows. In section 2 we
find the $U(1,1)$ and  $SL(2,R)$ symmetries for the wormhole scalar
model. In section 3, by choosing a factor ordering different from
the used in \cite{maha1,maha2} we show that the groups $U(2,1)$ and
$SL(2,R)$ are symmetries for the axion-dilaton string cosmology. In
section 4, for both models, we find some new solutions for the
Wheeler-DeWitt equation, including wave packets.  By imposing the
Hartle-Hawking boundary conditions, wormhole type-solutions are
found. Finally, in section 5, we give our concluding remarks.

\section{Symmetries for the Hawking and Page wormhole scalar model}

Hawking and Page considered the Wheeler-DeWitt equation for the
massless scalar field $\phi$ in a Friedmann-Robertson-Walker (FRW)
spacetime
\begin{equation}
{\mathcal H}\psi(a,\phi)=
\frac{1}{2}\left(\frac{1}{a}\frac{\partial}{\partial
a}a\frac{\partial}{\partial
a}-\frac{1}{a^2}\frac{\partial^2}{\partial \phi^2 }-a^2\right)
\psi(a,\phi)=0, \label{Haw}
\end{equation}
whose independent solutions are \cite{HP}
\begin{equation}
\psi(a,\phi)=J_{\pm i\frac{m}{2}}(ia^2/2)e^{im\phi}.
\label{deg1}
\end{equation}
Notice that there exists a solution for each integer $m$,
corresponding to the angular momentum eigenvalue: ${\cal
L}\psi(a,\phi)\equiv  -i\frac{\partial \psi(a,\phi)}{\partial
\phi}=m\psi(a,\phi)$. Thus, the Wheeler-DeWitt equation has an
infinite number of eigenstates.

We define the creation and annihilation operators
\begin{eqnarray}
&&a_0=\frac{1}{\sqrt{2}}\left( -\sinh{\phi}\left(a+\frac{\partial}{\partial a}\right)+\frac{\cosh{\phi}}{a}\frac{\partial}{\partial \phi}\right),\\
&&\bar{a}_0=\frac{1}{\sqrt{2}}\left(-\sinh{\phi}\left(a-\frac{\partial}{\partial a}\right)-\frac{\cosh{\phi}}{a}\frac{\partial}{\partial \phi}\right),\\
&&a_1=\frac{1}{\sqrt{2}}\left(\cosh{\phi}\left(a+\frac{\partial}{\partial a}\right)-\frac{\sinh{\phi}}{a}\frac{\partial}{\partial \phi}\right),\\
&&\bar{a}_1=\frac{1}{\sqrt{2}}\left(\cosh{\phi}\left(a-\frac{\partial}{\partial a}\right)+\frac{\sinh{\phi}}{a}\frac{\partial}{\partial \phi}\right),
\end{eqnarray}
which satisfy the commutation relations $\left[a_\mu, {\bar a}_\nu
\right]=G_{\mu \nu}=\hbox{diag}(-1, 1)$, $\mu,\nu=0,1$. By means
of the coordinate transformation $x=a \sinh{\phi}$ and $y=a
\cosh{\phi}$, these operators become
\begin{eqnarray}
&&a_0=\frac{1}{\sqrt{2}}\left(-x+\frac{\partial}{\partial x}\right), \hspace{3ex} \bar{a}_0=-\frac{1}{\sqrt{2}}\left(x+\frac{\partial}{\partial x}\right),\label{CA1}\\
&&a_1=\frac{1}{\sqrt{2}}\left(y+\frac{\partial}{\partial y}\right),\hspace{5ex}
\bar{a}_1=\frac{1}{\sqrt{2}}\left(y-\frac{\partial}{\partial y}\right).
\label{CA2}
\end{eqnarray}
The angular momentum operator is
\begin{eqnarray}
{\mathcal L}=-i\frac{\partial}{\partial
\phi}&=&-i\left(y\frac{\partial}{\partial
x}+x\frac{\partial}{\partial y}\right)\label{lz}\\
&=&-i(a_0\bar{a}_1-a_1\bar{a}_0),
\end{eqnarray}
and the Hamiltonian (\ref{Haw}) can be written as
\begin{eqnarray}
{\mathcal H}&=&\frac{1}{2}\left[\frac{\partial^2}{\partial
y^2}-\frac{\partial^2}{\partial x^2}-(y^2-x^2)\right]\\
&=&\bar{a}_0a_0-\bar{a}_1a_1-1.
\label{hamcar}
\end{eqnarray}

The creation and annihilation operators defined above allow to find
the $U(1,1)$ hidden symmetry generators $\bar{a}_0a_0$,
$\bar{a}_0a_1$,  $\bar{a}_1a_0$ and $\bar{a}_1a_1$, which commute
with the Hamiltonian operator ${\mathcal H}$. Since the group
$U(1,1)$ is equal to $SU(1,1)\times U(1)$ \cite{BARS}, we find that
the $U(1)$ generator is
\begin{align}
J_0&=-{\bar  a}_0a_0+{\bar  a}_1a_1\label{2dsim1}, \\
&=-a_0{\bar a}_0+a_1{\bar  a}_1-2, \\
&=-({\mathcal H}+1),
\end{align}
and the non-trivial $SU(1,1)$ traceless generators are
\begin{eqnarray}
&&J_{00}={\bar  a}_0a_0+\frac{1}{2}J_0,\hspace{3ex}J_{01}={\bar  a}_0a_1,\label{2dsim2}\\
&&J_{10}={\bar  a}_1a_0,\hspace{10ex}J_{11}={\bar
a}_1a_1-\frac{1}{2}J_0\label{2dsim3}.
\end{eqnarray}
These symmetry operators are such that $J_{00}=J_{11}$, $[J_0,{\mathcal H}]=0$,
$[J_0,J_{\mu \nu}]=0$ and $[J_{\mu \nu},{\mathcal H}]=0$.

If we consider the possibility of a matter-energy renormalization
by introducing an arbitrary constant \cite{8},  equation
(\ref{Haw}) can be rewritten in the following form
\begin{equation}
\frac{1}{2}\left(\frac{1}{a}\frac{\partial}{\partial
a}a\frac{\partial}{\partial
a}-\frac{1}{a^2}\frac{\partial^2}{\partial \phi^2 }-a^2-2E\right)\psi_{Em}(a,\phi)=0.\label{Non-zeroHaw}
\end{equation}
Notice this equation enforces us to introduce the energy $E$ to
label the wavefunction. Since the operators $J_{\mu \nu}$ commute
with the Hamiltonian then they do not change the energy $E$ but
the angular momentum quantum number $m$. Thus, these generators
transform the degenerate states corresponding to a given energy
between themselves, in particular those for the zero-energy
$E=0$.

Also we can define the set of operators
\begin{eqnarray}
&&K_+ \equiv \frac{1}{2}\left(-{\bar  a}_0^2+{\bar  a}_1^2\right),\label{2dk-1}\\
&&K_- \equiv \frac{1}{2}\left(-a_0^2+a_1^2\right),\label{2dk-2}\\
&&K_0 \equiv \frac{1}{2}\left(-{\bar  a}_0a_0+{\bar
a}_1a_1+1\right)=-\frac{{\mathcal H}}{2},\label{2dk-3}
\end{eqnarray}
which satisfy the commutation relations
\begin{eqnarray}
[K_+,K_-]=-2K_0,\hspace{5ex}
[K_0,K_{\pm}]=\pm K_{\pm}.
\end{eqnarray}
This means that the operators $K_0$, $K_+$ and $K_-$ close the
$SL(2,R)$ dynamical Lie algebra. A direct calculation shows that the
Casimir operator ${\hat K}\equiv K_0(K_0-1)-K_+K_-$  is related to
the angular momentum ${\mathcal L}$ as $K^2=-{\mathcal
L}^2-\frac{1}{4}$. From this result, the common eigenfunctions for
the Hamiltonian $K_0$ and the Casimir $K^2$ operators of the
$SL(2,R)$ algebra can be chosen as those of the Hamiltonian and the
angular momentum operators. Thus, from equation (\ref{2dk-3}) we get
$K_0|E\hspace{.5ex} m \rangle =-\frac{E}{2}|E \hspace{.5ex}
m\rangle$, and from the second commutation relation we show that
$K_0K_\pm|E\hspace{.5ex} m \rangle =-\left(\frac{E\mp
2}{2}\right)K_\pm |E \hspace{.5ex} m\rangle$. These results imply
that $K_\pm|E\hspace{.5ex} m \rangle \propto |E\mp 2 \hspace{.7ex}
m\rangle$. Hence, the operators (\ref{2dk-1}) and (\ref{2dk-2})
acting on the states $|E\hspace{.5ex} m \rangle$ change the energy
and leave fixed the angular momentum quantum number. If we restrict
the solutions to those of the Wheeler-DeWitt equation without
matter-energy renormalization, we must consider the states with
$E=0$, and the above $SL(2,R)$ dynamical symmetry is not relevant.

\section{Symmetries of axion-dilaton string cosmology}

We begin summarizing some important points of the Maharana papers
\cite{maha1,maha2} which are relevant to our work. In these
references it has been found the $SO(2,1)$ symmetry of axion-dilaton
string cosmology derived from the action in the Einstein frame
\begin{equation}
S=\int
d^4x\sqrt{-g}\left(R-\frac{1}{2}\partial_\mu\phi\partial^\mu\phi-\frac{1}{2}
e^{2\phi} \partial_\mu\chi\partial^\mu\chi\right),
\end{equation}
where $R$ is the scalar curvature, $\sqrt{-g}$ is the determinant of
the metric $g_{\mu\nu}$, and $\phi$ and $\chi$ are the dilaton and
axion fields, respectively. The homogeneous and isotropic FRW metric
for closed universes ($k=1$)
\begin{equation}
ds^2=-dt^2+a(t)^2\left(\frac{dr^2}{1-r^2}+r^2d\Omega^2\right),
\end{equation}
was assumed, where $a(t)$ is the scalar factor and $t$ is the cosmic
time. The corresponding Wheeler-DeWitt equation is
\begin{equation}
H\Psi:=\frac{1}{2} \left(\frac{\partial^2}{\partial a^2}
+\frac{p}{a}\frac{\partial}{\partial a}-a^2+\frac{1}{a^2}\hat
{C}\right)\Psi=0, \label{WDWMAHA}
\end{equation}
where in order to solve the ordering ambiguity between $a$ and
$\partial/\partial a$, it was  adopted the prescription $p=1$.
Since the action $S$ is invariant under the $SO(2,1)$
transformations (S-duality), also $H$ is invariant under these
transformations. $\hat C$ is the $SO(2,1)$ Casimir operator, which
expressed in the pseudospherical coordinate system
\begin{equation}
x=a \sinh \alpha \cos \beta,\hspace{2ex}y=a \sinh \alpha \sin
\beta, \hspace{2ex}z=a \cosh \alpha,\label{sphericalparabo}
\end{equation}
is just the Laplace-Beltrami operator given by
\begin{equation}
{\hat C}=-\frac{1}{\sinh{\alpha}}\frac{\partial}{\partial
\alpha}\left( \sinh{\alpha}\frac{\partial}{\partial \alpha}\right)
-\frac{1}{\sinh^2{\alpha}}\frac{\partial ^2}{\partial \beta^2}.
\label{angularmaha}
\end{equation}
The axion and dilaton fields can be written in terms of the
pseudospherical coordinates (\ref{sphericalparabo}) as
\begin{equation}
\chi=\frac{\sinh{\alpha}\cos{\beta}}{\cosh{\alpha}+\sinh{\alpha}\sin{\beta}},
\hspace{2ex}e^{-\phi}=\frac{1}{\cosh{\alpha}+\sinh{\alpha}\sin{\beta}}.
\end{equation}
The explicit solutions for the Wheeler-DeWitt constraint
(\ref{WDWMAHA}) on the pseudosphere were obtained from the
$SO(2,1)$ group theory by identifying that the correct series
involved in quantum cosmology is the continuous one \cite{maha2}.
These are
\begin{equation}
\Psi(a,\alpha,\beta)=J_{\pm
i\frac{\nu}{2}}(ia^2/2)Y^m_{-\frac{1}{2}+i\lambda}(\cosh{\alpha},\beta),
\label{deg2}
\end{equation}
where
$Y^m_{-\frac{1}{2}+i\lambda}(\cosh{\alpha},\beta)=e^{im\beta}P^m_{-\frac{1}{2}+i\lambda}(\cosh{\alpha})$,
and $\nu^2=(\lambda^2+\frac{1}{4})$, are the eigenfunctions for
the non-compact operator ${\hat C}$ and the compact generator
$-i\partial_{\beta}$, with
$P^m_{-\frac{1}{2}+i\lambda}(\cosh{\alpha})$ the associated
Legendre polynomials (also called toroidal functions).
Notice that for this case, by varying $\lambda$ and  $m$  there exists and
infinite degeneracy.

One of the main results of this paper is to find the full
symmetries for the Wheeler-DeWitt equation (\ref{WDWMAHA}). This
is based on recognizing that equation (\ref{WDWMAHA}) in the
coordinates (\ref{sphericalparabo}) with factor ordering $p=2$ can
be written as
\begin{equation}
H\Psi=\frac{1}{2}\left(\frac{\partial ^2}{\partial
z^2}-\frac{\partial ^2}{\partial y^2}-\frac{\partial ^2}{\partial
x^2}-(z^2-y^2-x^2)\right)\Psi=0. \label{WDWmahacart}
\end{equation}
Some aspects about the factor ordering operator in the context of
string cosmology are important to remark. When one considers the
Wheeler-DeWitt equation in the string frame the operator ordering
is usually fixed by T-duality invariance of the Hamiltonian. This
selection of the factor ordering is important in the graceful exit
problem of quantum string cosmology of pre-big bang scenario. In
the case of homogeneous and isotropic cosmology without the axion,
the T-duality is just the scale factor duality $a \rightarrow
\frac{1}{a}$ and this requirement constraint the choice of factor
ordering to $p=1$ \cite{GAS,GMV}. However, the scale factor
duality is not adequate in the graceful exit in pre-big bang
string cosmology when quantum loop corrections are taken into
account \cite{CCM}. Besides, the presence of an homogeneous  axion
field or spatial curvature is compatible with S-duality but breaks
T-duality (O(d,d) symmetry) \cite{CEW,MPD}. In our case, the use
of Einstein frame helps to show S-duality of the theory but it is
not useful to fix the factor ordering because $ a \rightarrow a$
under S-duality \cite{maha1}.

If we want to preserve reparametrization invariance of the
Hamiltonian, following the arguments presented in \cite{GAS}, we
need  $p=2$ because in our case the minisuperspace is
three-dimensional unlike to that found in \cite{HP},  where the
minisuperspace is bidimensional and therefore the adequate factor
ordering results to be  $p=1$.

For this model we propose the set of creation and annihilation
operators
\begin{eqnarray}
a_0&=&\frac{1}{\sqrt{2}}\left(-z+\frac{\partial}{\partial z}\right),\hspace{2ex}
\bar{a}_0=-\frac{1}{\sqrt{2}}\left(z+\frac{\partial}{\partial z}\right),\label{mahacre1}\\
a_1&=&\frac{1}{\sqrt{2}}\left(y+\frac{\partial}{\partial y}\right),\hspace{2ex}
\bar{a}_1=\frac{1}{\sqrt{2}}\left(y-\frac{\partial}{\partial y}\right),\label{mahacre2}\\
a_2&=&\frac{1}{\sqrt{2}}\left(x+\frac{\partial}{\partial x}\right),\hspace{2ex}
\bar{a}_2=\frac{1}{\sqrt{2}}\left(x-\frac{\partial}{\partial x}\right)\label{mahacre3}.
\end{eqnarray}
These operators satisfy the commutation relations $\left[a_\mu,
{\bar a}_\nu \right]=$ $G_{\mu \nu}=$ $\hbox{diag}(-1, 1,1)$, $\mu,\nu=0,1,2,$ and
factorize the Hamiltonian as follows
\begin{equation}
H=-{\bar a}_0a_0+{\bar a}_1a_1+{\bar a}_2a_2+\frac{3}{2}.
\end{equation}
The operators (\ref{mahacre1})-(\ref{mahacre3}) allow us  to define the
angular operators
\begin{eqnarray}
J_z&=&-i({\bar a}_1a_2-{\bar a}_2a_1)=-i\left(y\frac{\partial}{\partial x}-x\frac{\partial}{\partial y}\right),\\
J_y&=&-i({\bar a}_0a_1-{\bar a}_1a_0)=i\left(z\frac{\partial}{\partial y}+y\frac{\partial}{\partial z}\right),\\
J_x&=&-i({\bar a}_2a_0-{\bar a}_0a_2)=-i\left(x\frac{\partial}{\partial z}+z\frac{\partial}{\partial x}\right).
\end{eqnarray}
These operators satisfy the $SO(2,1)$ commutation relations
\begin{equation}
\left[J_x,J_y\right]=-iJ_z,\hspace{3ex}\left[J_z,J_x\right]=iJ_y,\hspace{3ex}\left[J_y,J_z\right]=iJ_x,
\end{equation}
and reproduce the Casimir operator  ${\hat C}=-J_z^2+J_y^2+J_x^2$.
This result reflects a non-compact symmetry (S-duality) on the angular part of
the Hamiltonians (\ref{WDWMAHA}) or (\ref{WDWmahacart}).

The creation and annihilation operators (\ref{mahacre1})-
(\ref{mahacre3}) allow to define the set of second-order
operators
\begin{eqnarray}
{\mathcal J}_{00}&=&{\bar a}_0a_0+\frac{1}{3}{\mathcal
J}_0,\hspace{5ex}{\mathcal J}_{11}={\bar a}_1a_1-\frac{1}{3}
{\mathcal J}_0 \label{second1}\\
{\mathcal J}_{22}&=&{\bar a}_2a_2-\frac{1}{3}{\mathcal
J}_0,\hspace{5ex}{\mathcal J}_{01}={\bar a}_0a_1,\label{second2}\\
{\mathcal J}_{10}&=&{\bar a}_1a_0\hspace{13ex}{\mathcal
J}_{02}={\bar a}_0a_2,\label{second3}\\
{\mathcal J}_{20}&=&{\bar a}_2a_0\hspace{13ex}{\mathcal
J}_{12}={\bar a}_1a_2,\label{second4}\\
{\mathcal J}_{21}&=&{\bar a}_2a_1\label{second5}\\
{\mathcal J}_0&=&-{\bar a}_0a_0+{\bar a}_1a_1+{\bar
a}_2a_2,\nonumber\\ &=&-a_0{\bar a}_0+a_1{\bar a}_1+a_2{\bar a}_2-3
\end{eqnarray}
Notice that the  operator ${\mathcal J}_0$ and the Wheeler-DeWitt
Hamiltonian (\ref{WDWmahacart}) are related by ${\mathcal
J}_0=H-\frac{3}{2}$. We prove that operators  ${\mathcal J}_0$ and
${\mathcal J}_{\mu \nu}$ satisfy the commutation relations
\begin{equation}
\left[{\mathcal J}_0,{\mathcal J}_{\mu \nu}\right]=0.
\end{equation}
This means that the operators ${\mathcal J}_{\mu \nu}$ are
the symmetries of the Hamiltonian $H$. In fact, the operators
${\mathcal J}_{\mu \nu}$ are the non trivial generators of
$SU(2,1)$, whereas ${\mathcal J}_0$ is the $U(1)$ generator
\cite{BARS}.

In a similar way to the wormhole model we can introduce a matter-energy
renormalization. Hence, the Wheeler-DeWitt equation (\ref{WDWMAHA}) takes the form
\begin{equation}
\frac{1}{2} \left(\frac{\partial^2}{\partial a^2}
+\frac{2}{a}\frac{\partial}{\partial a}-a^2+\frac{1}{a^2}\hat
{C}-2E\right)\Psi_{E\lambda m}=0. \label{ME}
\end{equation}
By using the spherical functions
$Y^m_{-\frac{1}{2}+i\lambda}(\cosh{\alpha},\beta)=e^{im\beta}P^m_{-\frac{1}{2}+i\lambda}(\cosh{\alpha})$
as the correct wavefunctions for the Casimir operator $\hat C$, we
propose $\Psi_{E \lambda m}$ to have the form
\begin{equation}
\langle a \alpha \beta|E \lambda m\rangle=e^{im\beta}P_{-\frac{1}{2}+i\lambda}^{m}(\cosh
\alpha)W_{E\lambda}(a).
\end{equation}
This allows us to find the solution to the Wheeler-DeWitt equation
with matter-energy renormalization for the scale factor
$W_{E\lambda}(a)$. It is given by
\begin{eqnarray}
W_{E\lambda}(a)&\equiv &
c_1a^{-\frac{3}{2}}M_{-\frac{E}{2},\Lambda}(a^2)\nonumber\\
&+&c_2a^{-\frac{3}{2}}W_{-\frac{E}{2},\Lambda}(a^2),
\end{eqnarray}
where $M$ and  $W$ are the Whittaker functions, and $\Lambda
\equiv \frac{\sqrt{2+4\lambda^2}}{4}$. Notice that to set the
correct coefficients, we need to impose on the functions
$W_{E\lambda}(a)$ one of the well known proposals for the boundary
conditions \cite{8,9,10,11,12}.

The toroidal functions  $P^{m}
_{-\frac{1}{2}+i\lambda} (x)$ \cite{JACKIW}, can be expressed in terms of the
hypergeometric functions \cite{maha2}
\begin{equation}
P^{m} _{j}
(x)=\frac{1}{\Gamma(1-m)}\left(\frac{x-1}{x+1}\right)^{\frac{m}{2}}{}_2F_1
\left( -j, j+1; 1-m;\frac{1-x}{2}\right),
\end{equation}
with $j=-\frac{1}{2}+i\lambda$. They satisfy the orthogonality
relations \cite{NVilenkin1}
\begin{equation}
\int_{1}^{\infty}P^{m} _{-\frac{1}{2}+i\lambda} (x)P^{m}
_{-\frac{1}{2}-i\lambda'} (x)dx = \delta(\lambda - \lambda'
)\Bigg|\frac{\Gamma(i\lambda)}{\Gamma\left( \frac{1}{2}+ i\lambda
-m \right)}\Bigg|^{2},
\end{equation}
and the completeness relation
\begin{equation}
\int_{0}^{\infty}P^{m} _{-\frac{1}{2}+i\lambda} (x)P^{m}
_{-\frac{1}{2}-i\lambda} (x')d\lambda = \delta(x - x')
\Bigg|\frac{\Gamma(i\lambda)}{\Gamma\left( \frac{1}{2}+ i\lambda
-m \right)}\Bigg|^{2}.
\end{equation}

Since operators $\mathcal{J_{\mu \nu}}$  commute with the
Hamiltonian $H$, it is immediate to show that the functions
$\mathcal{J_{\mu \nu}}|E \lambda m\rangle$ are also eigenfunctions
of the Wheeler-DeWitt equation (\ref{ME}). Taking into account
that the $\langle a \alpha \beta|E \lambda m\rangle$ are a
complete set of functions \cite{NVilenkin1}, we can use them to
expand  $\mathcal{J_{\mu \nu}}|E \lambda m\rangle$. Thus,
\begin{equation}
 \mathcal{J_{\mu\nu}}|E\lambda m \rangle= \sum_{m'=-\infty} ^{\infty} \int_{0}^{\infty}
C_{m,m'}(\lambda,\lambda')|E \lambda' m'\rangle d\lambda'\label{INT}.
\end{equation}
However, the analytical expression for the functions
$C_{m,m'}(\lambda,\lambda')$ is very difficult to obtain due to
the integrand involves both toroidal and Whittaker functions.

We define the new set of operators
\begin{eqnarray}
&&{\mathcal K}_+=\frac{1}{2}\left(-{\bar a}_0^2+{\bar a}_1^2+{\bar a}_2^2\right),\\
&&{\mathcal K}_-=\frac{1}{2}\left(-{a}_0^2+{ a}_1^2+{a}_2^2\right),\\
&&{\mathcal K}_0=\frac{1}{2}\left(-{\bar a}_0a_0+{\bar a}_1a_1+{\bar a}_2a_2+\frac{3}{2}\right)=\frac{H}{2},
\end{eqnarray}
which satisfy the $SL(2,R)$ commutation relations
\begin{eqnarray}
&&\left[{\mathcal K}_+,{\mathcal K}_-\right]=-2{\mathcal K}_0,\\
&&\left[{\mathcal K}_0,{\mathcal K}_\pm\right]=\pm {\mathcal K}_\pm.
\end{eqnarray}
The Casimir operator $\hat {\mathcal K}_0$ for this $SL(2,R)$
algebra and the Casimir operator $\hat C$ for the  $SO(2,1)$ angular
momentum are related by ${\hat {\mathcal K}}_0=-\frac{1}{4}{\hat
C}-\frac{3}{16}$. These results allow to find the action of the
operators $K_-$ and $K_+$ on the non-zero energy eigenstates of the
Wheeler-DeWitt equation,
\begin{equation}
{\mathcal K}_{\mp} |E\hspace{.5ex}\lambda \hspace{.5ex} m
\rangle\propto |E\mp 2 \hspace{1ex}\lambda \hspace{.5ex} m \rangle.
\end{equation}
Thus, if we restrict to the solutions of the
Wheeler-DeWitt equation (\ref{WDWmahacart}) with $E=0$, the
generators ${\mathcal K}_+$, ${\mathcal K}_-$ and ${\mathcal K}_0$
of the $SL(2,R)$ algebra do not play a relevant role as a symmetry
group for the Wheeler-DeWitt equation. However, the $SU(2,1)$
symmetry generators are relevant for any energy $E$, and describe  the
degeneracy in the quantum numbers $\lambda$ and $m$.

\section{Wave packets solutions for the scalar and axion-dilaton
string cosmology models}

In a similar way to that followed to calculate wave packets for the
Kantowski-Sachs spacetime \cite{Cava}, we can construct wave packets
solutions by integrating over the quantum number $m$
\begin{equation}
\psi_{WDW}=\int_{-\infty}^{\infty}e^{i(m\phi+\gamma)} J_{\pm
i\frac{m}{2}}(ia^2/2)dm.
\end{equation}
This allows us to obtain
\begin{equation}
\psi_{HP}^{wp}=A_{1,2}e^{\pm\frac{1}{2}a^2\cosh(2\phi+\gamma_{1,2})},\label{HPpaq}\\
\end{equation}
where $A$ and $\gamma$ are constants. These functions are solutions
for the Wheeler-Dewitt equation (\ref{Haw}) for the Hawking-Page
scalar model. The solution with the minus sign is the only which is
a gaussian wave packet and satisfy the Hartle-Hawking boundary
condition ("no-boundary proposal") \cite{8}, i. e. the wave function
of the universe is regular at $a\rightarrow 0$ and it is
exponentially damped for large scale factor $a\rightarrow \infty$.
Thus, the gaussian wave packet represents a wormhole solution for
the Wheeler-Dewitt equation \cite{HP}.

For the string cosmology model, we find the wave packets
\begin{equation}
\Psi_{SC}^{wp}={\mathcal A}_{\pm}e^{\pm
\frac{1}{2}a^2\cosh(2\alpha+\gamma_{\pm})},\label{bue}
\end{equation}
which are solutions for the Wheeler-Dewitt equations $(H\pm
\frac{1}{2})\Psi=0$.  These are completely analogous to those for
the Hawking-Page scalar model (\ref{HPpaq}). Therefore, the solution
with the minus sign corresponds to a whormhole type-solutions for
the axion-dilaton string cosmology model.

Also, we can show that the functions
\begin{equation}
\Psi_{SC}=e^{\pm\frac{1}{2}(z^2+y^2)}\sqrt{x}\left(C_1I_{\frac{1}{4}}(
x^2/2)+C_2K_{\frac{1}{4}}(x^2/2)\right),\label{buenas}
\end{equation}
are solutions for the Wheeler-DeWitt equation $H\Psi=0$, being
$I_{\frac{1}{4}}( x^2/2)$ and $K_{\frac{1}{4}}(x^2/2)$ the modified
Bessel functions and $C_1$ and $C_2$ constants. In these solutions
if we interchange the $x$- and $y$-coordinates, the resulting
functions are also  solutions for the Wheeler-DeWitt equation.
However, only the solution with the minus sign and $C_2=0$ is
regular at the origin (this is because the only modified Bessel
function which leads to a regular wave function as $a\rightarrow 0$
is $I_{\frac{1}{4}}(x^2/2)$). Thus, the functions
\begin{equation}
\Psi_1=Ce^{-\frac{1}{2}(z^2+y^2)}\sqrt{x}I_{\frac{1}{4}}(
x^2/2)\label{bue1}
\end{equation}
and
\begin{equation}
\Psi_2=Ce^{-\frac{1}{2}(z^2+x^2)}\sqrt{y}I_{\frac{1}{4}}(
y^2/2)\label{bue2}
\end{equation}
are wormhole type-solutions for the axion-dilaton string cosmology
model. To our knowledge the solutions above in this section do not
have been reported in the literature.

From the point of view of the group theory, it is more important the
$U(1)$ generator than the dynamical equation (in this case, the
Wheeler-Dewitt equation). For the Hawking-Page and the axion-dilaton
string cosmology models, we find the following solutions
\begin{equation}
\psi_{U(1)}^{hp}=e^{\pm\frac{1}{2}a^2},\hspace{8ex}
\Psi_{U(1)}^{st}=e^{\pm \frac{1}{2}a^2}\label{paq2},
\end{equation}
which satisfy the Wheeler-DeWitt equations $({\mathcal H}\mp
1)\psi=0$ and $(H\mp \frac{3}{2})\Psi=0$, respectively.  In fact,
these solutions are annihilated by either of the two forms for the
$U(1)$ generators $J_0$ and ${\mathcal J}_0$, respectively. These
are manifestly Lorentz invariant under rotations around the $z$-axis
and under the $SO(2,1)$ group, respectively. However, from the
quantum cosmology point of view, these solutions do not represent
any interesting scenario because they do not involve the scalar
field ($\phi$) or the axion-dilaton ($\chi-\phi$) fields.\\

\section{Concluding remarks}

In this paper we have found the symmetries related to the
Wheeler-DeWitt Hamiltonian for the Hawking and Page wormhole and the
axion-dilaton quantum cosmology. We have shown that the
Wheeler-DeWitt Hamiltonian for the wormhole model has the $U(1,1)$
non-compact symmetry which describe the degeneracy of the states with or without
energy-matter renormalization. Also we have found that the Wheeler-DeWitt Hamiltonian
accept an additional $SL(2,R)$
dynamical symmetry when energy-matter renormalization is considered.
In this case, we showed that the $SL(2,R)$ dynamical symmetry generators transform
the states from an energy Hilbert eigensubspace to another.

The factor ordering is frequently chosen by convenience \cite{COULE,KONTO,LIDSEY1,LIDSEY2}.
For the axion-dilaton string
cosmology we have set the factor ordering $p=2$, which is necessary
to preserve reparametrization invariance of the Hamiltonian. Indeed,
this choice was crucial in order to find
the closed Lie algebras representing the hidden symmetries of the
axion-dilaton string cosmology. A similar setting has been taken
by Pioline, {\it et. al.} \cite{PIOLINE} in order to fix the
conformal symmetry $SO(2,1)$ for the one-dimensional
Wheeler-DeWitt equation.
For the axion-dilaton string cosmology
Hamiltonian, the non-compact hidden symmetries $U(2,1)$ and
$SL(2,R)$ are permissible. The $U(2,1)$ symmetry is valid whenever energy-matter
renormalization is or not present.
Also, in this case, the $SL(2,R)$ hidden symmetry transforms the states from
an energy Hilbert eigensubspace to another.

The zero-energy eigenfunctions (\ref{deg1})
and (\ref{deg2}) are particular cases of the huge degeneracy on the states $\psi_{Em}$  or  $\Psi_{E\lambda m}$
when matter-energy renormalization is considered.
The huge degeneracy of the wave functions on the secondary quantum numbers  ${m}$ or ${\lambda m}$ for the systems studied in
this paper  are fully described by the non-compact symmetries $U(1,1)$ or $U(2,1)$, respectively.  On the other
hand, the $SL(2,R)$ symmetries are suitable to relate the states
with different principal quantum number but maintaining the
secondary quantum numbers fixed.

Other symmetries have been found for the Wheeler-DeWitt equation in
different physical settings. For example, the $SO(2,1)$ conformal
group has been found for the one-dimensional radial Wheeler-DeWitt
equation with cosmological constant \cite{PIOLINE}. We emphasize
that the symmetries for the Wheeler-DeWitt equation found in this
work are for the complete Hamiltonian.

Wormhole type-solution for the Hawking-Page model, equation
(\ref{HPpaq}), and for the axion-dilaton string cosmology, equations
(\ref{bue})-(\ref{bue2}), were found. To our knowledge these
solutions do not have been reported in previous works.

Finally, our procedure can be applied to find the symmetries of the
Wheeler-DeWitt equation for other systems like multidimensional
quantum wormholes \cite{HAW1,ZHUK}, or the Kantowski-Sachs quantum
cosmological model \cite{KANTO,LIDSEY3}, which is work in
progress.

\section{Acknowledgments}

This work was partially supported by SNI-M\'exico, CONACYT grant
number J1-60621-I, COFAA-IPN, EDI-IPN, SIP-IPN projects numbers
20100897, 20110127 and  20100684

\end{document}